\documentclass[12pt,a4paper]{article}
\usepackage[utf8x]{inputenc}
\usepackage[T2A]{fontenc}
\usepackage[ukrainian,english]{babel}
\usepackage{titling}
\usepackage{caption2}
\usepackage{ucs}
\usepackage{amsmath}
\usepackage{amsfonts}
\usepackage{amssymb}
\usepackage{graphicx}

\newcommand{\textsup}[1]{$^{#1}$}

\newcommand{\Hh}{$^{\rm h}$}
\newcommand{\Mm}{$^{\rm m}$}
\newcommand{\Ss}{$^{\rm s}$}
\newcommand{\Dd}{$^{\rm d}$}
\newcommand{\fm}{\mbox{$.\!^{\rm m}$}}
\newcommand{\fd}{\mbox{$.\!^{\rm d}$}}

\begin{document}
\DeclareRobustCommand{\authorEng}{\begin{tabular}{c}
	Natalia A. Virnina$^{1}$, Ivan L. Andronov$^{1}$, Maxim V. Mogorean$^{2}$ \\[0.8em]
    $^{1}$ \parbox[t]{0.9\textwidth}{\small Department “High and Applied Mathematics”, Odessa National Maritime University, Odessa, Ukraine, \nolinebreak{virnina@gmail.com}, \nolinebreak{ilandronov@gmail.com} }\\[0.5em]
    $^{2}$ \parbox[t]{0.9\textwidth}{\small Mariinskaya Grammar School, Odessa, Ukraine, \nolinebreak{maximmogorean@gmail.com} }\\[0.5em]
\end{tabular} }

\title{A Hot Spot and Mass Transfer in the Algol-type Binary System WZ Crv}
\author{\authorEng}
\date{}
\maketitle

\renewcommand{\abstractname}{ABSTRACT.}
\begin{abstract}
We present the results of the first two color VR observations of the Algol-type binary system WZ Crv 
(12 \textsup{h}44\textsup{m}15.19\textsup{s}, -21\textsup{d}25\textsup{m}35.4\textsup{s} (2000)) 
which were obtained using the remotely controlled telescope TOA-150 of Tzec Maun Observatory. We determined the moments of individual minima, the orbital period and its derivative, the initial epoch, color indices V-R and temperature estimates of the components. Also we noticed that the phase curve is asymmetric: the second maximum is higher than the first one. It indicates that there is a spot in the photosphere of one of the stars in this system.
\end{abstract}



\vspace{2em}

\section{INTRODUCTION.}
The Algol-type binary system WZ Crv had been discovered in 1936 and published by Luyten \cite{c1} in 1937 with the number 686. Then it got the preliminary numbers NSV 05914, EC 12416-2109 and GSC 6110.00930. The primary minimum is rather deep. According to Luyten the brightness range is from 13\fm0 to 15\fm0.
In the General Catalog of Variable Stars (Samus’ et al. \cite{c2}) the value of the period of WZ Crv is $P=1\fd78878,$ and the range of the brightness variations is 12\fm76 – 14\fm7 in V-band.
During the Edinburg-Cape Blue Object Survey (Kilkenny et al. \cite{c3}) the spectral classes G and F7 for the components had been determined.
Two minima timings were observed by Otero \& Dubovsky \cite{c4} (CCD observations) and Locher \cite{c5} (visual observations). 

\section{OBSERVATIONS}. 
We observed the south Algol-type binary system WZ Crv (12\Hh{}44\Mm{}15.19\Ss{}, -21\Dd{}25\Mm{}35.4\Ss{}) = NSV 5914 during the period from JD 2455321 to JD 2455395 using the remotely controlled astrophysical refractor TOA-150 of Tzec Maun Observatory ($D=150$mm, $F=1100$mm), located in Moorook, South Australia. This telescope was equipped with SBIG ST-L-6K CCD camera with a filter wheel. The field of view is 57.5’ x 86.3’, the pixel scale is 3.37"/pixel. We observed in photometric V-Bessel and R-Bessel filters and got 317 fit points in each filter. The 120 sec exposures were used. The images were calibrated by dark frames and flat field frames.

To test our photometric system we observed the field of the variable star AE Aqr. The BVRI magnitudes in standard bands of 228 stars in the vicinity of AE Aqr were determined by Henden \cite{c6}. We used 93 most bright stars and determined the color transformation coefficients:
\begin{align*} 
&\frac{d(R_i-R)}{d(V-R)} = 0.015\pm0.035;\\
&\frac{d(V_i-V)}{d(V-R)} = -0.033\pm0.043.
\end{align*}  
where $R_i$ and $V_i$ are the magnitudes in the instrumental system, and $V$ and $R$ are the standard
magnitudes (by Henden). This means that our photometry system is close to the standard one.

There is no photometry data for reference stars in the field of WZ Crv in AAVSO
database or in any previous papers about this star. But there are SDSS \cite{c7} photometry data in the
“u, g, r, i, z” photometric system for the stars in the vicinity of WZ Crv. We used 3 reference
stars to calibrate our photometry: USNO-A2.0 0675-12226641 (12\Hh{}44\Mm{}54.624\Ss{}, -21\Dd{}30\Mm{}27.21\Ss{}),
USNO-A2.0 0675-12232415 (12\Hh{}45\Mm{}35.274\Ss{}, -21\Dd{}33\Mm{}42.93\Ss{}), USNO-A2.0 0675-12233718 (12\Hh{}45\Mm{}44.211\Ss{}, -21\Dd{}16\Mm{}07.06\Ss{}). The transformation formulae (Lupton, 2005, \cite{c8}) were used to convert the “u, g, r, i, z” data into BVR system. The information about the reference stars is given in the Table \ref{reference_stars_table}. Their positions and the position of the investigated star WZ Crv are shown on the Fig. \ref{reference_stars_chart}.

\begin{center}
\captionof{table}{}\label{reference_stars_table}
\flushleft \footnotesize
\begin{tabular}{|*{10}{c|}}
\hline
\bf \# & \bf USNO-A2.0 & \bf u  & \bf g  & \bf r  & \bf i  & \bf z  & \bf B  & \bf V  & \bf R  \\\hline
\bf 1  & 0675-12226641 & 15.219 & 13.870 & 13.295 & 13.230 & 13.340 & 14.277 & 13.534 & 13.132 \\\hline
\bf 2  & 0675-12232415 & 15.781 & 13.953 & 13.248 & 13.504 & 12.818 & 14.401 & 13.541 & 13.179 \\\hline
\bf 3  & 0675-12233718 & 16.534 & 14.369 & 13.532 & 13.284 & 13.046 & 14.858 & 13.881 & 13.315 \\\hline
\end{tabular}

\end{center}

\begin{center}
\centering
\includegraphics[width=\textwidth]{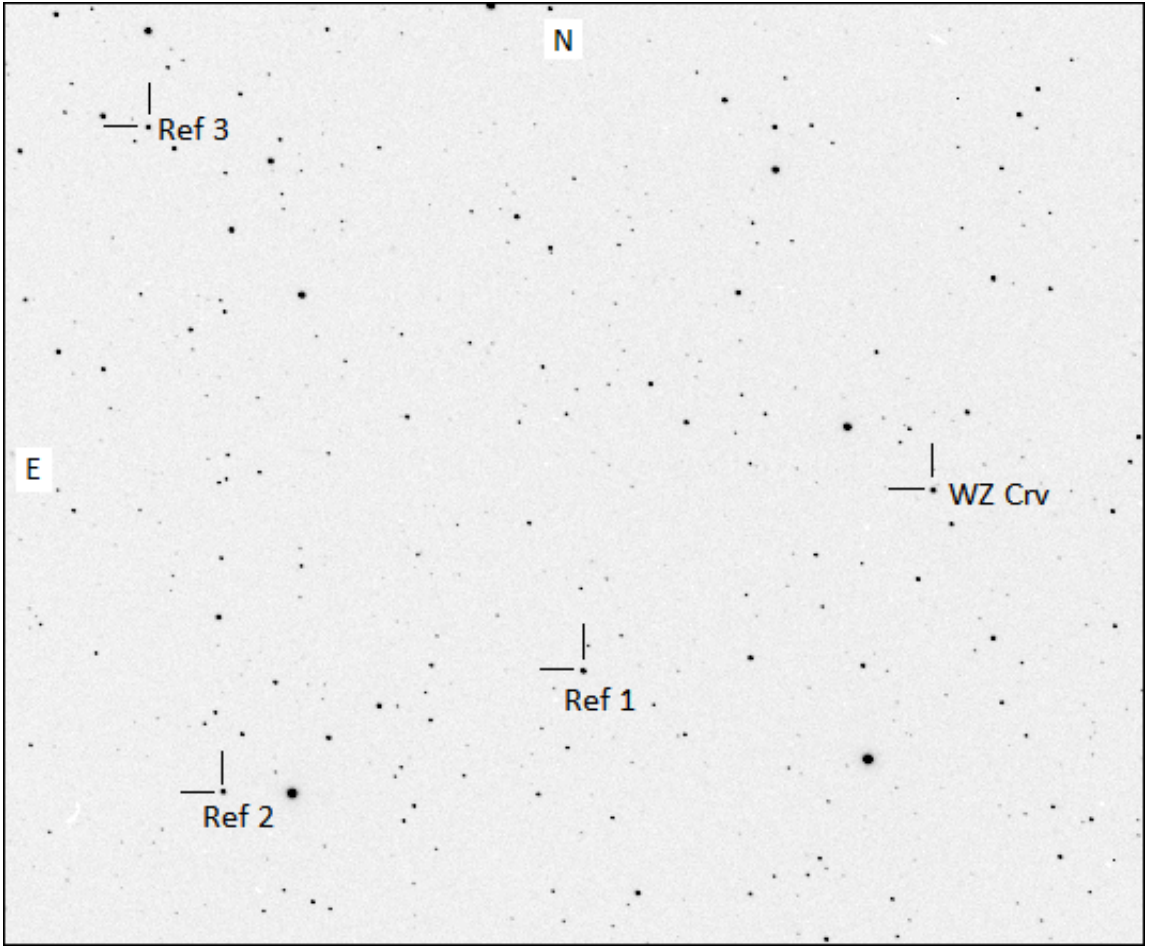}
\captionof{figure}{Finding chart. The field of view is 30’x25’. The positions if three reference stars and WZ Crv are marked. North is up, East is left.}\label{reference_stars_chart}
\end{center}

\section{Data analyzing.}
We determined the period, the initial epoch and the values of maxima and minima in
both, V and R, bands with corresponding errors using our observations. The software FDCN
(Andronov \cite{c9,c10}) was used to compute the coefficients of the statistically optimal
trigonometric polynomial fit, using the least squares method, which allows to calculate the
necessary parameters and their errors estimates. The statistically optimal degree of the
trigonometric polynomial $s=9$. The determined values are:
\begin{align*}
P&= 1\fd788788 \pm 0\fd000033\\
T_0&={\rm HJD} 2455354.63500 \pm 0.00031\\
{\rm min}_{I}(R)&=14\fm004 \pm 0\fm005		& {\rm min}_{I}(V)&=14\fm564 \pm 0\fm009 \\
{\rm min}_{II}(R)&=12\fm751 \pm 0\fm006	& {\rm min}_{II}(V)&=13\fm052 \pm 0\fm010 \\
{\rm Max}_{I}(R)&=12\fm588 \pm 0\fm005		& {\rm Max}_{I}(V)&=12\fm889 \pm 0\fm006 \\
{\rm Max}_{II}(R)&=12\fm534 \pm 0\fm005	& {\rm Max}_{II}(V)&=12\fm868 \pm 0\fm007. \\
\end{align*}

The phase curves are shown on the Fig. \ref{figure_phase_curves_own}. Besides, we made the V-R phase curve; the
result is presented on the same Figure in the same scale. It shows that in the primary minimum
the temperature is lowest, and in the phase of the secondary minimum the temperature is highest.

\section{Analyzing of the Phase Curves}
At the main (deeper) minimum, there is a horizontal part which is typically interpreted as
a total eclipse of a smaller star (called “primary”) by a larger one (called “secondary”), so during
this time interval we see only a larger star. The brightness of the star is $m_2=m_{\rm min\,I}$. To determine
this value, we have computed a sample mean value of the measurements during this phase
interval of assumed constant brightness either in the filter $V$, or $R$. To determine the brightness
of the eclipsed star, we have to use the expression $E_0=E_1+E_2$, where $E_0$, $E_1$ and $E_2$ are intensities
of the binary system (both stars), primary and secondary stars, respectively. The intensities $E$ are
related to brightness (stellar magnitudes) $m$ using the classical Pogson’s formulae:
$$ \frac{E}{E_S}=10^{-0.4(m-m_S)}, $$
$$ m-m_S=-2.5\lg\frac{E}{E_S}, $$
where $E_S$ and $m_S$ are the intensity and brightness of some calibration source. From these
expressions, one may determine the magnitude of the eclipsed primary star as
$$ m_1 = m_0 - 2.5\lg{\left(1-10^{-0.4(m_2-m_0)}\right)} $$
The value of $m_0$ is a subject of a special discussion. Sometimes the out-of-eclipse parts of the
light curve are not constant due to the “reflection” effect (heating of atmosphere of one star by an
emission of another star) and the “ellipticity” effect (due to a tidal distortion of one or both
stars). From the mathematical point of view, these effects may be modeled as a second-order
trigonometric polynomial
$$ x(\phi)=C_1 + C_2\cos(2\pi\phi)+C_3\cos(4\pi\phi)+C_4\sin(2\pi\phi)+C_5\sin(4\pi\phi). $$
The coefficients $C_4$ and $C_5$ are close to zero for symmetric light curves. However, for WZ Crv,
the maximum at phase 0.75 is brighter than that at the opposite phase 0.25, so $C_4$ is distinctly
different from zero, whereas $C_5=0$ within error estimates. Thus we decreased the number of the
determined parameters to 4. The formulae for the error estimates were presented by e.g.
Andronov \cite{c9,c10}. The shape of the secondary minimum shows a linear-like descending and
ascending branch linked with a parabola. Thus we have used the "asymptotic parabola" fit
(Marsakova {\&} Andronov \cite{c11}). Results for the brightness maxima in both filters $V, R$ are
presented above. We estimated the smoothed brightness $m_0$ at the phase $\phi=0$ of $V=12\fm953 \pm 0\fm007$
and $R=12\fm634 \pm 0\fm004$. For $\phi=0.5$, $V=12\fm917 \pm 0\fm006$ and $R=12\fm601 \pm 0\fm004$. Thus the brightness of the primary is $V_1=13\fm233 \pm 0\fm014$, $R_1=13\fm000 \pm 0\fm008$ and $(V-R)_1=0\fm237 \pm 0\fm016$. Similarly, we determined the brightness of the eclipsed part of the secondary corresponding to the phase $\phi=0.5$: $V_{2e}=15\fm251 \pm 0\fm014$, $R_{2e}=14\fm824 \pm 0\fm054$ and $(V-R)_{2e}=0\fm427 \pm 0\fm109$. The ratio of the intensities from the secondary to the eclipsing part is $E_2/E_{2e}=1.94 \pm 0.20$ and $E_2/E_{2e}=2.19 \pm 0.05$ for V and R. These values are the same within their error estimates. If interpreting this ratio as the ratio of the eclipsed surfaces (in the simplest model of zero limb darkening), one may estimate the ratio of the radii $R_2/R_1=(E_2/E_{2e})^{1/2}$ of $1.40 \pm 0.07$ and $1.48 \pm 0.04$, respectively. Theoretical achievements are that this ratio is constant. However, the slight difference between the ratio $E_2/E_{2e}$ for $V$ and $R$ may be interpreted as a larger coefficient of the limb darkening in $V$ as compared with that in $R$. This is in an agreement with results of Al-Naimiy \cite{c12}.

Using the color indices $V-R$, we estimated the temperatures of the primary and secondary
stars in this binary system. In the primary minimum the cooler bigger star eclipses the hotter star.
The flatness of the primary minimum (Fig. \ref{figure_shape}) indicates that in the primary minimum the eclipse is total. Thus, it is possible to determine the approximate temperature of the first component,
using the color index $V-R$ in the phase of primary minimum. 
We determined the corresponding temperature, using the empirical formula (Andronov {\&} Baklanov \cite{c13}), which describes the dependence between the color index $V-R$ and the spectral class. There is a misprint in the English version of the paper, the correct version is:
$$ \frac{10000}{T} = 1.948(\pm 0.011)+1.474(\pm 0.019)\cdot (V - R - 0.681). $$

In the total primary eclipse the value of the color index is $V-R=0\fm560±0\fm012$. Thus, the
temperature of the second component is $T = (5650 \pm 66)K$. Using the calibration of spectral
class temperatures of Allen (1973) \cite{c14} the “3-$\sigma$” temperature “corridor” corresponds to the
range of spectral classes of G2 – G6. Using the determined color index $(V-R)_1=0\fm237\pm0\fm016$ for
the primary component, we estimated the temperature of first component $(7874\pm 166)K$ and
range of spectral classes A6 – F1.

Assuming that the inclination of the orbit is rather close to $90^\circ$, we determined the lower
limits for the relative radiuses of the stars, using the formulae
$$ \frac{R_1}{a} = \frac{\sin(\pi D) - \sin(\pi d)}{2} $$
$$ \frac{R_2}{a} = \frac{\sin(\pi D) + \sin(\pi d)}{2} $$
where $R_1$ and $R_2$ are the average radiuses of the stars, a is the orbital separation, i. e. a distance between the centers of the components, $d$ is the duration of the total eclipse (in phases), $D$ is the total duration of eclipse, from the beginning till the very end (in phases too). The phases of the
eclipse are from $-0.0915$ to $0.0915$, the phases of the total eclipse are from $-0.017$ to $0.017$,
which means that the values of $D$ and $d$ are:
\begin{align*}
D&=0.183 \\
d&=0.034
\end{align*}

Thus, the lower limits of the relative radiuses:
\begin{align*}
\frac{R_1}{a}&= 0.219\\[0.25em]
\frac{R_2}{a}&= 0.325
\end{align*}
The ratio $R_2/R_1=1.48$ determined from geometrical constrains is an excellent agreement with that
obtained using the fluxes. Thus we conclude that the inclination angle is close to $90^\circ$.

The most interesting feature of the phase curves is the statistically significant difference
between values of maxima in both filters. This difference reaches $0\fm060 \pm 0\fm003$ in filter $R$
and $0\fm039 \pm 0\fm005$ in filter $V$. The cause of such effect might be a cold or a hot spot in the
photosphere of one of the stars.

To find out if the spot is steady, we checked the ASAS database. There are 493 fit points
in filter V, which were obtained during the period from JD2451876 to JD2455049. Using these
observations we determined the period and the initial epoch. The values are
$P=1\fd788797 \pm 0\fd0000021$ and $T_0=HJD2453494.2755 \pm 0.0014$ respectively. The phase curve is
shown on the Fig. \ref{figure_curves_asas}. There is no statistically significant asymmetry on this curve, perhaps, because of rather big error of individual observations. Thus, using ASAS data, we can’t conclude on the stability of the spot.

On the other hand, according to the previous calculations, the spectral classes of the
components of WZ Crv are G2 – G6 and A6 – F1, which nearly completely excludes the
possibility of forming rather steady big cold spots in the photosphere of at least one of the stars.
More possible solution is the presenting of a hot spot as the result of deep impact of the accretion
stream from the first component (donor) directly into the photosphere of the second one
(accretor) without forming the accretion disk. In this case the hot spot will be rather steady,
which might be confirmed by further observations.

\section{Period Variations and Mass Transfer}
To study stability of the orbital period using the O-C analysis \cite{c15}, we compiled a list of
the minima timings. The individual times of minima were published by Otero \& Dubovsky
(2004) and Locher (2005). added the median time of minima, calculated from our observations,
from the NSVS database and from the ASAS database. ASAS data covers rather long period of
observations (9 years), so we divided this array into three parts (JD 2451876-2452879, JD
2452977-2453908, JD 2454091-2455049) and analyzed them individually. For the ASAS data,
we have used several methods. The period and the initial epoch of the best "complete"
trigonometric polynomial fit of order 9 were used for further analysis. As for these data there is
no evidence for a statistical difference of maxima, we have used a “symmetric” trigonometric
polynomial fit
$$ x(\phi) = C_1 + \sum_{j=1}^S C_{j+1}\cos(2\pi j (\phi-\phi_0)), $$
where $\phi$ is the phase corresponding to the moment of time $t$: $\phi=(t-T_0)/P-INT((t-T_0)/P)$, where $T_0$ is an “initial epoch”, $P$ is a period and $INT(x)$ is an integer part of x.

The coefficients $C_j$ were determined using a linear method of least squares and the
parameter $\phi_0$ (phase of the main minimum) was determined by iterations (non-linear method of
least squares). Using the value of $\phi_0$, we have computed a “mean” moments of time of minimum, which are closest to the mean time of the observations similar to a “complete” trigonometric
polynomial fit $x(\phi)=C_1 + \sum_{j=1}^S C_{j+1}\cos(2\pi j(\phi-\phi_{0j}))$ with phases $\phi_{0j}$, which are individual for each harmonic. Symmetric trigonometric polynomial of 9-th order were used for the NSVS data and the 3 intervals of the ASAS. They are listed in the Table \ref{table_minima_timings}.

\begin{center}
\vbox{\includegraphics[width=\textwidth]{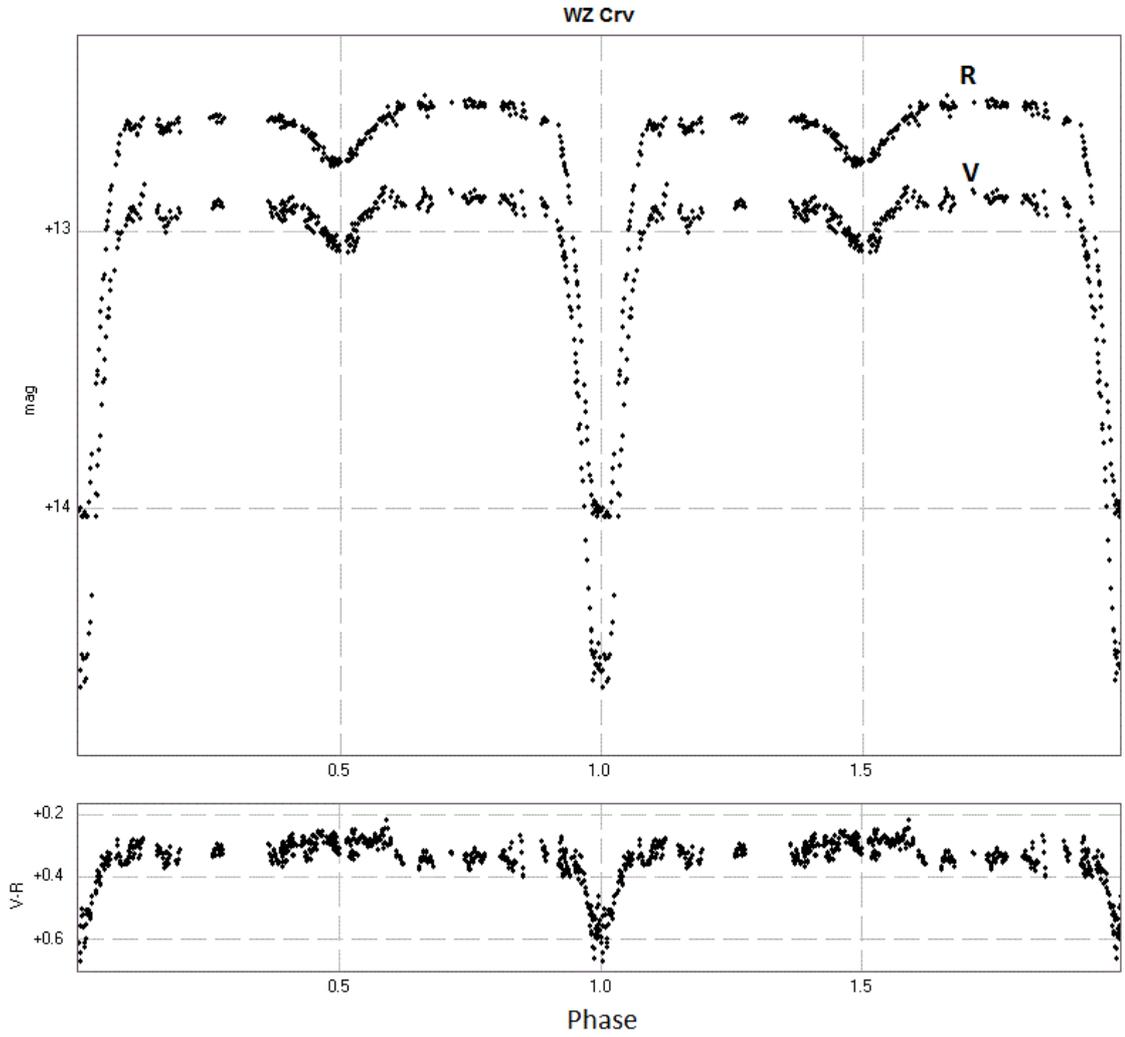}
\captionof{figure}{The phase curves of WZ Crv in R and V filters and the phase curve of the color index.}
\label{figure_phase_curves_own}}
\end{center}

\begin{center}
\vbox{\includegraphics[width=\textwidth]{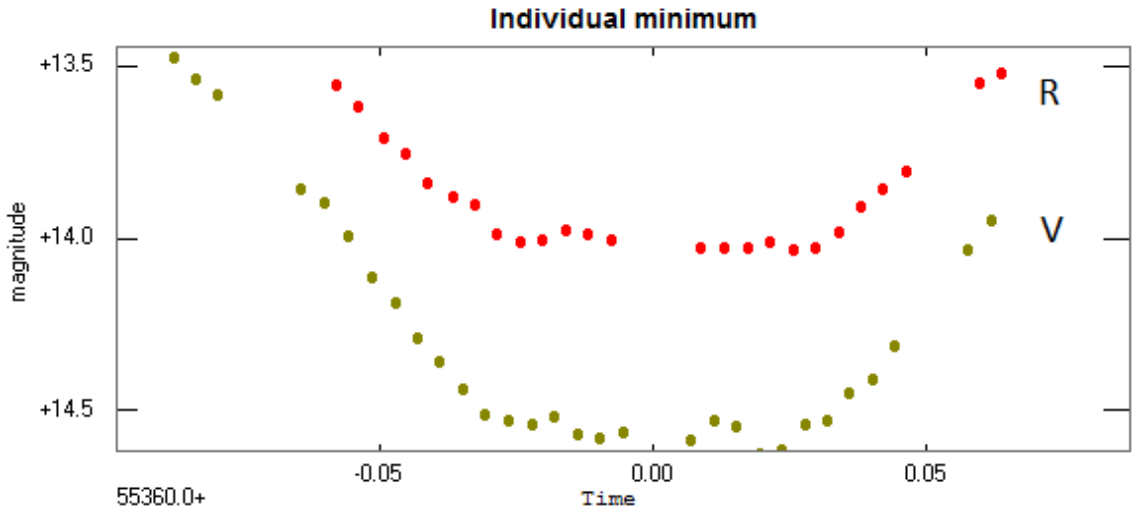}
\captionof{figure}{The shape of the primary minimum in the phase of full eclipse in R and V filters, obtained on JD 2455360 (2010.06.12).}
\label{figure_shape}}
\end{center}

\begin{center}
\vbox{\includegraphics[width=\textwidth]{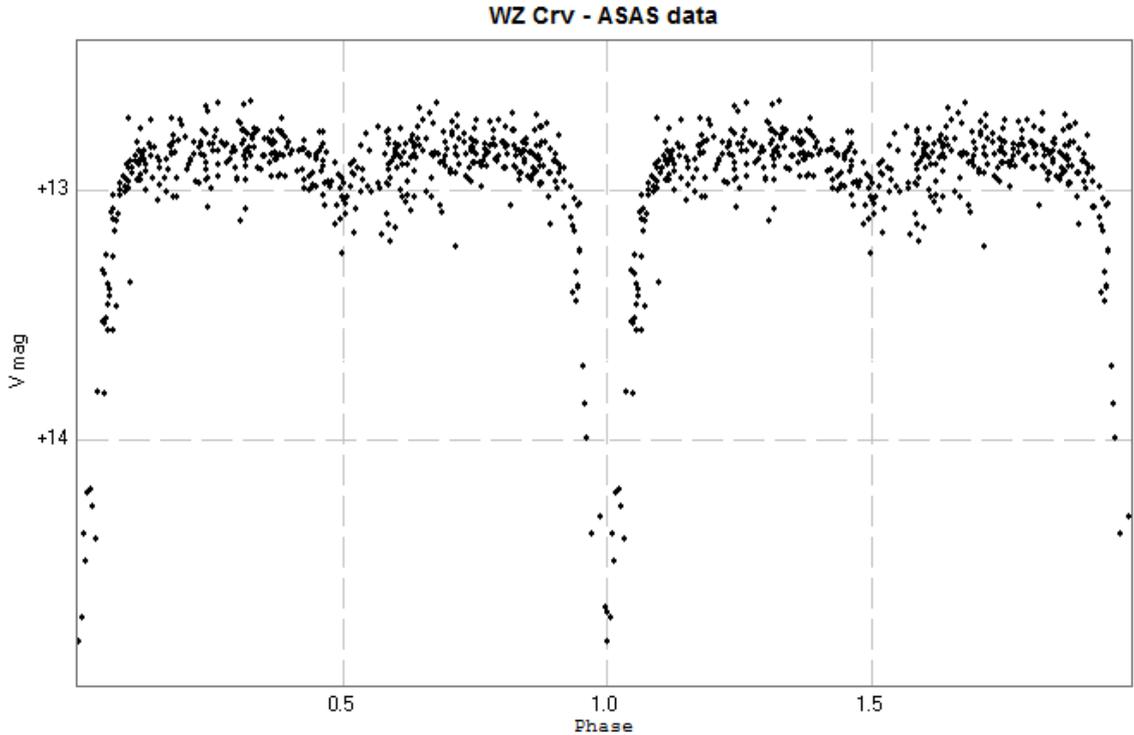}
\captionof{figure}{WZ Crv phase curve according to ASAS data in V-band.}
\label{figure_curves_asas}}
\end{center}

\begin{center}
\vbox{
\captionof{table}{Minima timings.}
\begin{tabular}{|c|c|r|r|l|} \hline
    HJD       & HJD error &   E  &    O-C   & source   \\\hline
2451596.37218 & 0.00103   & 0    &  0       & NSVS     \\\hline
2452086.49800 & 0.00800   & 274  & -0.00209 & Otero    \\\hline
2453440.59300 & ~	  	  & 1031 & -0.01961 & Locher   \\\hline
2452378.07042 & 0.00176   & 437  & -0.00212 & ASAS 1   \\\hline
2453483.53860 & 0.00201   & 1055 & -0.00492 & ASAS 2   \\\hline
2454535.36234 & 0.00242   & 1643 &  0.01148 & ASAS 3   \\\hline
2455354.63500 & 0.00031   & 2101 &  0.01923 & Virnina  \\\hline
\end{tabular}
\label{table_minima_timings}
}
\end{center}

For all ASAS data, we obtained $HJD_{min}=2453494.27648 \pm 0.00115$. However, this value
corresponds to a wide interval and thus shifted from the smoothing curve. The minimum of
Locher deviates significantly from the fit as well. This is due to a large error estimate, which is
by a factor of 3-27 times larger than the minima timings determined from our observations and
that of the NSVS and ASAS. Thus we excluded the values from the ASAS (three subintervals
together) and of Locher \cite{c5} from our analysis. As there is no error estimate of the timing was
published by Otero {\&} Dubovsky \cite{c4}, we have used the same weight for all minima timings.

Thus, we analyzed 7 times of minima, and made the O-C diagram as the dependence
between the cycle number and the residual $O-C=T_E-(T_0+PE)$, i.e. the difference between the
observed and calculated times of minima. The O-C diagram exhibits a parabolic shape, which is
characteristic for secular period variations (cf. Tsessevich \cite{c16}). The smoothing parabola is
shown on the Fig. \ref{figure_o_c}. The least-squared method yields the equation for time of minima:
$$
T_{min} = \mathop{2451596.3722}_{\pm 0.002} + 
		  \mathop{1.788795}_{\pm 0.000002} \cdot (E-918) + 
		  \mathop{1.09 \cdot 10^{-8}}_{\pm 0.31 \cdot 10^{-8}} \cdot (E-918)^2 
$$

From the diagram and the equation it is clear that the orbital period and, according to
Kepler’s law, the distance between the components increases. In assumption of conservative
system, we determined $dP/dt$:
$$
	dP/dt=(1.2\pm 0.3)\cdot 10^{-8} {\rm days/yr},
$$

and the characteristic time of the period variations is
$$
	P/(dP/dt)=(4.02\pm 1.13) \cdot 10^5 {\rm yr}.
$$

It means that the matter streams rather fast from the less massive star to the more massive one. In
the case of WZ Crv it means that the second component lost its mass and the matter streams
throw the inner Lagrange point falls into the Roche lobe of the second component, which
completely consistent with the direct impact accretion stream version of a hot spot.

\begin{center}
\vbox{\includegraphics[width=\textwidth]{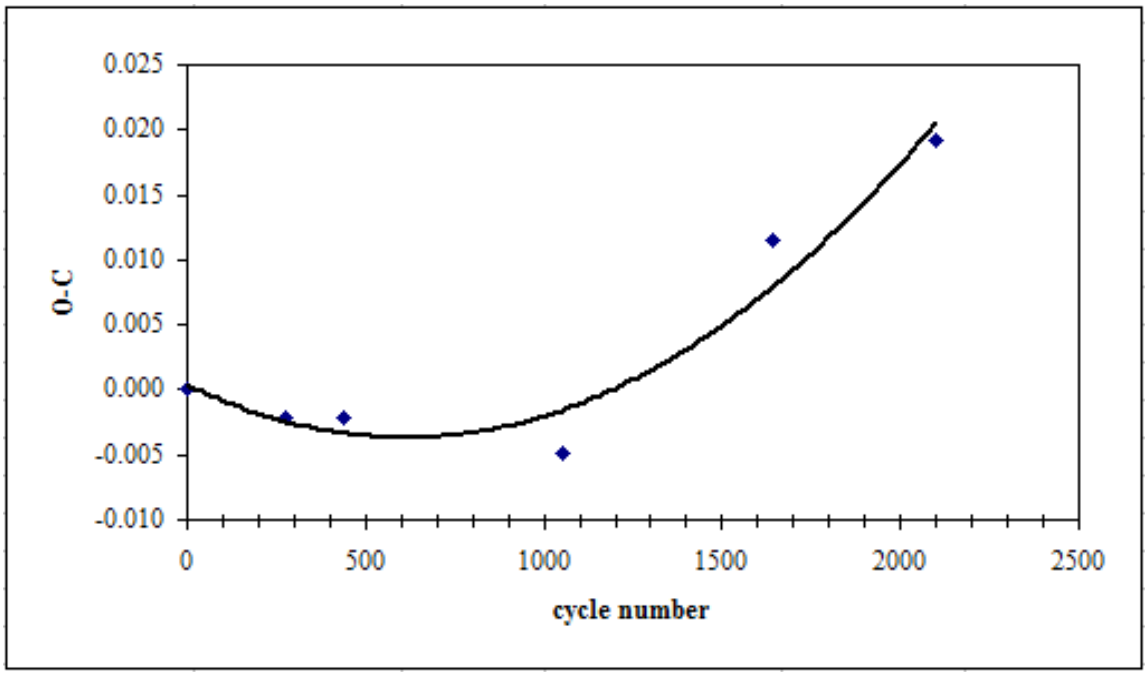}
\captionof{figure}{O-C diagram.}
\label{figure_o_c}}
\end{center}

\section{Conclusions}
\begin{itemize}
\item The poorly studied Algol-type eclipsing binary system WZ Crv was investigated for the
first time in two photometric systems $V$ and $R$.
\item The brightness and color indices were determined at the primary and secondary minima
and maxima.
\item The brightness at maxima differs, indicating a possible spot in the atmosphere of one of
the stars.
\item The temperatures, spectral classes and radii of both components were estimated.
\item The increase of the orbital period was discovered.
\end{itemize}

\section{Acknowledgements}
This investigation was based on data collected using the telescope of the Tzec Maun
Observatory, operated by the Tzec Maun Foundation. The authors are very thankful to Ron
Wodaski (director of the observatory) and Donna Brown-Wodaski (director of the Tzec Maun
Foundation).

\include{titleUkr}
\end{document}